\documentclass[conference]{IEEEtran}

\usepackage[utf8]{inputenc}
\usepackage[hidelinks = true]{hyperref}

\usepackage{url}
\usepackage{graphicx}
\usepackage{acronym}
\usepackage{xcolor}

\usepackage{footnote}

\IEEEoverridecommandlockouts

\begin{document}

\author{
    \IEEEauthorblockN{Samuel Gomes}
	\IEEEauthorblockA{
		\textit{INESC-ID \& Instituto Superior Técnico,}\\
		\textit{University of Lisbon, Portugal}\\
		samuel.gomes@tecnico.ulisboa.pt
	}
	\\[0.3cm]
	\IEEEauthorblockN{João Dias}
	\IEEEauthorblockA{
		\textit{INESC-ID \& Faculty of Science and Technology,}\\
		\textit{University of Algarve, Portugal}\\
		joao.dias@gaips.inesc-id.pt
	}
	\\
	\and
	\IEEEauthorblockN{Tomás Alves}
	\IEEEauthorblockA{
		\textit{INESC-ID \& Instituto Superior Técnico,}\\
		\textit{University of Lisbon, Portugal}\\
		tomas.alves@tecnico.ulisboa.pt
	}
	\\[0.3cm]
	\IEEEauthorblockN{Carlos Martinho}
	\IEEEauthorblockA{
		\textit{INESC-ID \& Instituto Superior Técnico,}\\
		\textit{University of Lisbon, Portugal}\\
		carlos.martinho@tecnico.ulisboa.pt
	}
   
}

\title{Reward-Mediated Individual and Altruistic Behavior\\
\thanks{This work was supported by national funds through Fundação para a Ciência e a Tecnologia (FCT) with references SFRH/BD/144798/2019, SFRH/BD/143460/2019, and UIDB/50021/2020.}
}
\maketitle


\acrodef{Model}[BRM]{Behavior-Reward Model}

\begin{abstract}
Recent research has taken particular interest in observing the dynamics between altruistic and individual behavior. This is a commonly approached problem when reasoning about social dilemmas, which have a plethora of real world counterparts in the fields of education, health and economics. 
Weighing how incentives influence in-game behavior, our study examines individual and altruistic interactions, by analyzing the players' strategies and interaction motives when facing different reward attribution strategies.
Consequently, a model for interaction motives is also proposed, with the premise that the motives for interactions can be defined as a continuous space, ranging from self-oriented (associated to self-improvement behaviors) to others-oriented (associated to extreme altruism behaviors) motives.
To evaluate the promotion of individual and altruistic behavior, we leverage Message Across, an \textit{in-loco} two-player videogame with adaptable reward attribution systems.
We conducted several user tests (N = 66) to verify to what extent individual and altruistic reward attribution systems led players to vary their strategies and motives orientation.
Our results indicate that players' strategies and self-reported orientation of interaction motives varied highly significantly upon the deployment of individual and altruistic reward systems, which leads us to believe on the suitability of applying an incentive-based strategy to moderate the emergence of individual and altruistic behavior in games.
\end{abstract}


\acresetall

\section{Introduction}
\label{section:Introduction}

Since the last century, researchers have studied the dynamics between individual and altruistic behavior \cite{hamilton1963evolution,kurman2006self}. This is the most important target of analysis in social dilemma scenarios \cite{hilbe2014cooperation}. For instance, in public goods games \cite{hilbe2014cooperation, du2017promotion, parks1994social,galbiati2008obligations, correia2019exploring}, subjects have to decide between contributing with their own investments to individual or social goods, thus deciding between developing self and others' welfare.
Along social dilemmas, this self-vs-others paradigm is also predominant in education. Namely, the theory of Self-determination \cite{ryan2000self, richter2015studying, cruz2017need} distinguishes between several levels of motivation, namely intrinsic motivation - related to self definition of goals and satisfaction from self development; and extrinsic motivation - influenced by external factors like rewards, approval or competition.

Over the years, research on behavior promotion has taken particular interest in the use of games to change players' long-term commitments \cite{tanenbaum2013three,schuller2013serious,chow2019can}, for instance to develop environment sustainability awareness \cite{Lee2015,Ouariachi2019} or non-sedentary behaviors \cite{Arundell2019,Hagood2016}.
As an interactive medium, games allow the promotion of feelings of competence through feedback and rewards, and support relatedness through social interactions such as competition and cooperation \cite{richter2015studying}.
Therefore, we credit that promoting in-game behavior can be a useful path to approach aspects of attitude change such as the motives which drive interactions, given the growing impact of games in players' lives \cite{chatfield2010fun}.
This line of research can help to inspire the parameterization of systems aiming to balance or enhance individual and altruistic facets of behavior among people with different individual characteristics or cultural backgrounds, as there is evidence that these intrinsic characteristics are able to drive different game strategies \cite{parks1994social}. 
Researchers have focused on studying how 
the attribution of different rewards (ranging from simple scores to collectibles, resources, item granting systems, achievement systems, feedback messages, etc.) \cite{cruz2017need, wang2011game} affect player experience.
In fact, although some recent studies, detailed further on, embraced reward-based behavior promotion \cite{siu2014collaboration, mcclintock1966reward}, there is the consensus that further research in the motives upon interpersonal choice behavior is still needed \cite{mcclintock1966reward, Vegt2016}.
Altogether, this work contemplates the following general research question:
\\
\\
\textit{How can rewards be used to mediate individual and altruistic behavior?}
\\
\\
To answer this problem, we leverage Message Across, an \textit{in-loco} two-player word matching game, in which we implemented two versions of the score attribution system, aimed at orienting the players' interactions to either themselves or others.
We then conducted a user testing phase where pairs of participants played the different versions without knowing what score systems were being deployed at each moment. We extracted the players strategies and scores through the course of the game, as well as their self-reported orientation of interaction motives (between self-oriented and others-oriented motives) to find answers for our research question.

The remaining of the paper is organized as follows: in Sections~\ref{section:RelatedWorkI} and~\ref{section:RelatedWorkII}, we explore what interaction styles, as well as ways to promote in-game behavior were identified so far by related research, and can be analyzed to model and moderate individual and altruistic interactions;
In Section~\ref{section:SolutionDescription}, we describe how we implemented individual and altruistic score systems in Message Across; 
Next, we include the evaluation process in Section \ref{section:ExperimentalSetup}, and present and discuss the empirical results in Section~\ref{section:Results}; 
Finally, we summarize the work and finish with future directions in Section \ref{section:Conclusions}.

\section{From Theory to a Model for Interaction Motives}
\label{section:RelatedWorkI}

We started our analysis by observing what behavior mediation techniques were identified in social dilemmas scenarios.
A significant amount of research deploying social dilemmas focuses on collaborative interactions, studying how can higher levels of cooperation and altruism be fostered. 
For instance, Du and Gerla presented a mobile social networks model, and verified, through multi-agent simulations, that cooperation was promoted due to the degree heterogeneity and regular moving patterns \cite{du2017promotion}.
Correia et. al \cite{correia2019exploring} examined prosociality through the social attributes and responsibility attribution levels reported by subjects playing an iterated public goods game with uncertainty, alongside two robots. The game strategies of each robot were manipulated so that one focused exclusively on the public welfare - the cooperator, while the other focused exclusively on individual welfare - the defector. The outcome of the game was also manipulated, namely whether the subjects survived through the game or the public good financially collapsed. Results indicated that the prosocial robot was rated more positively in terms of its social attributes than the selfish robot, regardless of the game result, and that when players lost the game, they tended to attribute significantly more responsibility to the defector robot. 
This trend may indicate that interactions with individual motives might be more easily discovered or remembered than interactions with altruistic motives, namely whenever a bad outcome is encountered.
Hilbe et al. \cite{hilbe2014cooperation} theoretically examined several strategies to sustain cooperation in a public goods game and volunteer social dilemmas, including generalized variants of Tit-for-Tat and Win-Stay Lose-Shift. More importantly, to define such strategies, three particular sub-classes of strategies were identified: the fair-neutral strategy which simply ensures individual payoff is aligned with the average group payoff; the extortionate strategy which, in scenarios where mutual defection leads to the lowest group payoff, is based on ensuring that individual payoffs are above average (related to free-riding and individual focus, not concerning for the others); and the generous strategy which, in scenarios where mutual cooperation is the social optimum, consists in letting co-players gain higher payoffs, in order to develop social welfare.
We believe that strategies such as extortionate and generous are useful to promote when allowing mixed interactions, as they follow two opposite poles. In one pole, there is an individual motive for interaction, without attention for others, and in the other pole there is an others-oriented motive for interaction, without valuing self consequences.
Following this line of thought, we further analyzed theories regarding these two opposite ideas, leading us to further define a model for interaction motives.

Hoping to exclusively study individual motives for interaction, we analyzed several work regarding self-improvement.
Self-Improvement can be defined as a conscious desire to improve self ability \cite{sedikides2009self}, a result of self evaluation. Numerous research focused on what can influence this behavior, and how it emerges in general task-based and education-oriented scenarios \cite{sedikides2009self,kurman2006self,o2017does,dunn1999deliberate}. 
Kurman et al. \cite{kurman2006self} indicate two major subprocesses of self-improvement: the recognition stage, which consists in identifying faults or to recognize the need for improvement, and the action stage which consists in actively implementing an idea, that is, to take action.  
In fact, self-enhancement -- a pervasive motivation or goal directed force that manifests broadly to promote and protect the positivity of the self --, is claimed to hinder the recognition phase, but facilitate the action phase and overall task performance, by which it is argued that self-enhancement is related to self-improvement \cite{kurman2006self, o2017does}.
Task-based scenarios as games can foster self-improvement, as task-related self-enhancement seems to effectively facilitate task performance, opposed to non task-related self-enhancement \cite{o2017does}.
Following this line of thought, we believe that some multiplayer games can foster self-improvement through their actions, by directing players to embrace choices which individually improve themselves, without directly taking into consideration the actions of other players.

In order to study the others-oriented motives for interaction, we examined multiple theories related to altruism. 
Seelig and Rosof present several categorizations and review several research regarding altruism \cite{seelig2001normal}, from which we highlight Kitayama's scale of altruism as masochism \cite{kitayama1991wounded}.
On a study related to the dual nature of the feminine ideal in Japanese culture, the author defines altruism as a continuum between two facets.
While the first facet can be interpreted as an interaction consisting of mutual help between peers, the second facet happens when people feel that it is important to do good for others, even if it means the process will not be pleasant for them. In other words, people engaging in the second type of altruism are \textit{exclusively} motivated by others, without even trying to minimize the negative consequences that helping others might bring to themselves.
Although some research associates this effect to \textit{Psychic Altruism}, we will refer to this interaction style as \textit{Extreme Altruism}.

\subsection{Model of Interaction Motives}

\begin{figure}
	\centering
	\includegraphics[width=1.0\linewidth]{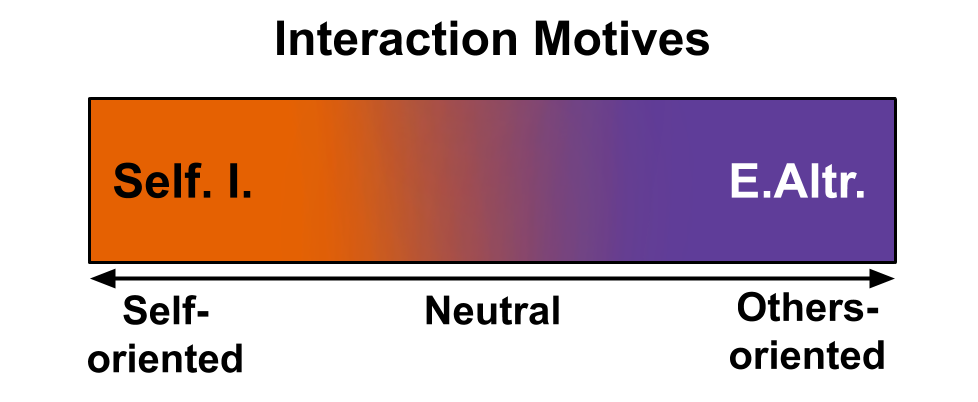}
	\caption{\label{fig:motivesSpace} Continuous space organizing interaction motives, between Self-oriented and Others-oriented. Self-oriented motives can be associated to the Self Improvement behavior, while Others-oriented motives can be associated to the Extreme Altruism behavior.}
\end{figure}

The presented related research approached several interaction styles, from which two extremes were identified: \textit{Self-Improvement} and \textit{Extreme Altruism}. 
In this work we deploy both these ends of the spectrum, placing them as extreme poles of a continuous interactions motive space (see Fig.\ref{fig:motivesSpace}). This categorization allows us to examine to what extent players interactions are individually or altruistically motivated.
In one pole, we have an individual motive for interaction, without attention for others, and in the other pole there is an others-oriented motive for interaction, without valuing self consequences.
Given this interaction motives model, the presented related research led us to deem that adequate indicators to measure the emergence of individual and altruistic behavior (besides extracting and analyzing the actual game actions and scores) was to acquire self-reported motives for interaction, comprehended between \textbf{Self-oriented} and \textbf{Others-oriented}. 

\section{Mediating Individual and Altruistic Behavior}
\label{section:RelatedWorkII}

In order to verify how to mediate individual and altruistic behavior, we also examined work specifically devoted to behavior promotion.
For instance, Vegt et al. \cite{Vegt2016} 
showed that different game rules could generate distinct reported player experiences and observable distinct player behaviors, which could be further discriminated into four patterns: expected patterns of helping and ignoring, and unexpected patterns of agreeing and obstructing.
Returning to social dilemma scenarios, Rosen and Haaga \cite{rosen1998facilitating} managed to induce higher levels of cooperation in small groups of four to eight subjects (higher number of cooperative social dilemma game actions, and higher altruistic attitudes toward a specific dilemma story), by applying message-based persuasion methods. In particular, the authors defined a social dilemma differently to participants in different conditions. 
To promote cooperation, instead of neutrally defining the nature of the dilemma, two explanations were applied: either the positive effects of collaboration and negative effects of free-riding were directly exposed for the considered social dilemma problem; or subjects were told that the same task was presented to various professionals at a conflict-resolution conference, who agreed that \textit{cooperation was the only appropriate response to the conflict and was necessary for societal harmony in general}.
Galbiati and Vertova, on the other hand, studied the promotion of cooperation through contribution obligations (minimum to which players have to contribute) in a public goods game \cite{galbiati2008obligations}. After performing several tests with no, low and high valued obligations, the authors concluded that high obligations led players to, in average, focus more in cooperating, while not reducing as much their cooperative contributions through the course of the game. 
Moreover, when the obligations unexpectedly increased, the players re-increased their cooperation levels. 
Besides obligations, research has studied several gameplay-oriented reward attribution approaches, ranging from simple scores to collectibles, resources, item granting systems, achievement systems, player dossiers, and feedback messages \cite{cruz2017need,wang2011game,richter2015studying,medler2011player}.
In particular, rewards in the form of score attribution systems are considered simple and appropriate mechanisms for comparisons between players, as they can be easily presented and understood \cite{richter2015studying,wang2011game,mcclintock1966reward}. They also have the advantage of not affecting gameplay, which provides more flexibility when parameterizing behavior. Based on this thought, we decided to follow this simple route for behavior promotion.

\section{Solution Description}
\label{section:SolutionDescription}

\begin{figure}[!htbp]
	\centering
	\includegraphics[width=1.0\linewidth]{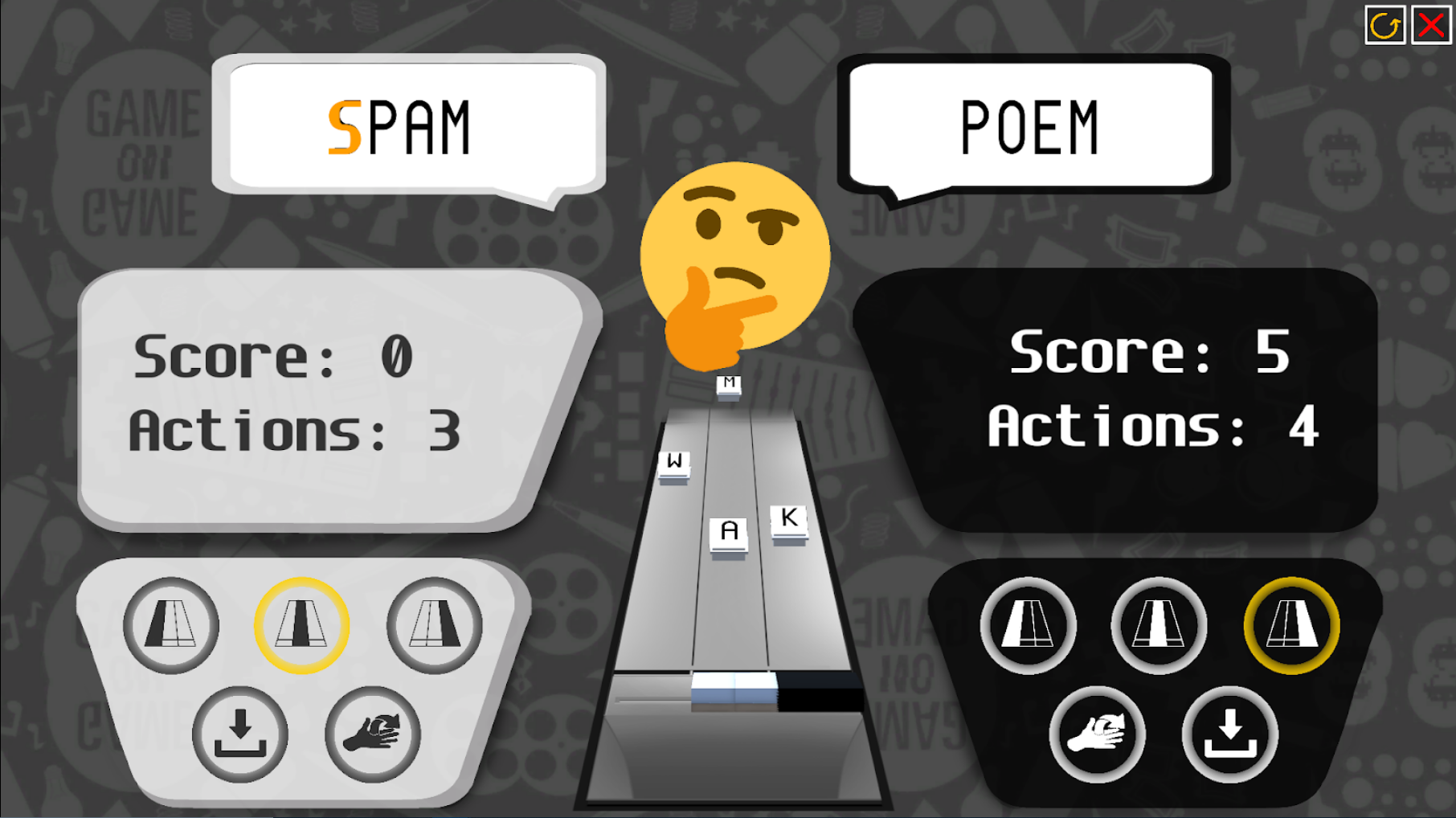}
	\caption{\label{fig:messageAcrossScreenshot} Screenshot of the game Message Across.}
\end{figure}

As previously commented, we used a game called Message Across\footnote{The implementation of the game is available online, hosted in the platform \textit{GitHub}: \url{https://github.com/SamGomes/message-across}} to examine the dynamics of players' interactions. A screenshot of the game is provided in Fig.~\ref{fig:messageAcrossScreenshot}. In the course of the game, players try to complete words as they advance through the levels. Each level presents two words on the top of the screen, one for each player. 
Because we wanted to promote interactions which led players to vary their strategies and interaction motives, and because we wanted words which were easily completed, only four letter words with two letters in common were considered in our experiments.
In the middle of the screen, the game presents a track containing three lanes where letters move towards players. The track also contains two markers, one for each player, arranged at the bottom.

In order to select a letter, a player has to move his/her marker to the lane where the letter is sliding, and select an action.
When the letter collides with the marker, the selected action is performed.
If two players are in the same track, only the first player that selects an action is able to perform it.

Players can perform one of two possible actions at each moment.
They can either \textbf{take the letter} or \textbf{give the letter} to the other player.
The objective of each player is to \textbf{obtain the highest score}. 
In our experiments, each player could perform a maximum of four actions per level, and a level finished whenever both players had no actions left to perform.
We believe that a limit of four possible actions exacerbated the players' need to search for meaningful strategies.

When reviewing related research, we examined the differences between individual and altruistic interactions, and built a model for interaction motives bounded between self-improvement and extreme altruism. To foster these two extreme behaviors in-game, we developed two divergent scoring versions:

\begin{itemize}
	\item
	The \textbf{Self-Improvement} version exclusively rewarded 10 points to players who took letters that were useful for them;

	\item
	The \textbf{Extremely Altruistic} version exclusively rewarded 10 points to players who gave letters needed by the other players.
\end{itemize}

The duality between these strategies is apparent, as the first rewards players which act for their own welfare, disregarding the actions of other players, while the second one rewards players who act exclusively for the others' welfare, disregarding the consequences of those actions for themselves.
Given that these reward versions were developed to allow players to understand the game in different manners, we predicted two distinct trends, which we translated to two hypotheses. Firstly, the players would embrace different, opposite strategies. These strategies would allow the game to motivate the players' interactions the towards opposite poles defined in our interaction motives model (Section \ref{section:RelatedWorkI}):
\\
\\
\textit{H1: The Self-Improvement version will implicitly drive players to perform a high number of takes and the Extreme Altruism version will drive players to perform a high number of gives.}
\\
\\
\textit{H2: The Self-Improvement version will implicitly drive players to report self-oriented interaction motives, and the Extreme Altruism version will implicitly drive players to report others-oriented interaction motives.}
\\
\\
In the next section, we present how we evaluated our approach considering these premises.

\section{Experimental Setup}
\label{section:ExperimentalSetup}

In order to evaluate the effectiveness of our score attribution strategies in driving in-game individual and altruistic interactions, we performed several experiments where pairs of participants played through the different versions of Message Across. In each test, the name of the game versions was obfuscated by using the letters to represent them, and the order of presentation of the game versions was uniformly randomized between groups to avoid any potential learning effects. \textbf{Therefore, throughout the session, the participants did not know how the game was being scored and had to figure that out by themselves}.


A touch screen was included in the experiment room for players to interact with the game. Two computers were used for our experiments: one computer executed the game and other computer allowed players to self-report their interaction motives orientation through a questionnaire. A Go-Pro video camera was also included for observing player movements and in-game activity and to help remedy some possible inconsistencies in automatic data collection. The camera was attached to a tripod and positioned approximately 50 cm in front of the touch screen.
The camera view of the our setup can be observed in Fig. \ref{fig:experimentalSetup}.

\subsection{Sample}
Participants were recruited, in teams of two, through standard convenience sampling procedures including direct contact and word of mouth.
Subjects included anyone interested in participating if they were at least 18 years old.
There were no potential risks and no anticipated benefits to individual participants. 
We conducted a total of 37 tests in a college laboratory.
Participation was open to outside visitors, which meant that not all participants were college students.
After data analysis, four tests did not meet quality criteria, e.g in-game data not recorded or questionnaires with missing answers.
Thus, our final data set comprised 33 tests, a total of 66 participants (37 males, 26 females) between 18 and 40 years old \((M = 23.12; SD = 4.09)\).

\subsection{Procedure}
\begin{figure}[!htbp]
	\centering
	\includegraphics[width=0.9\linewidth]{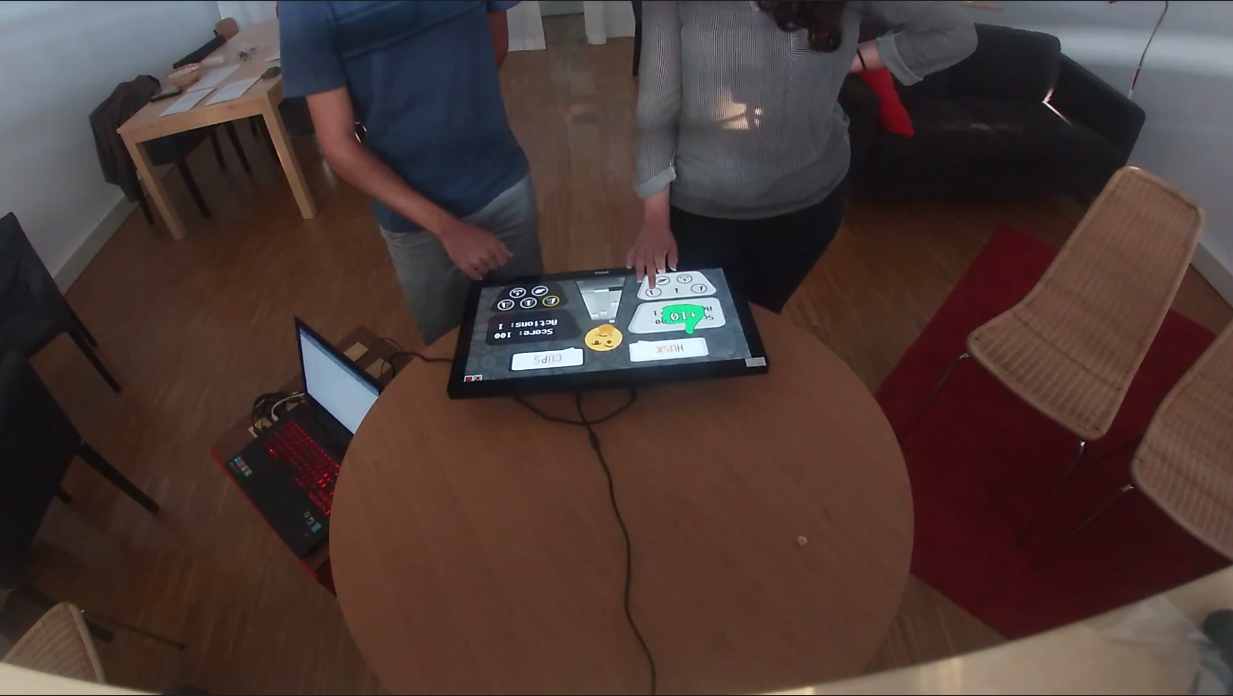}
	\caption{\label{fig:experimentalSetup} Camera view of a group playing Message Across during our experiments.}
\end{figure}

The experiment operated as follows:
(i) Firstly, participants were informed about the experience and invited to sign a mandatory consent form. They were also informed that they could stop the experiment at any time;
(ii) After signing the consent form, both participants were asked to move next to the touchscreen (as seen in Figure~\ref{fig:experimentalSetup}) and received a tutorial regarding in-game mechanics and possible actions to perform.
Additionally, participants were allowed to play up to seven levels without being rewarded for any \textit{give} or \textit{take} action, in order to support the development of fluent playing skills;
(iii) When both participants felt comfortable with the game mechanics, they played the two game versions in random order with each gaming session lasting seven levels.
After each gaming session, participants were asked to complete questionnaires measuring their interaction motives orientation regarding that session.
At the end of the experiment, participants received a candy bar as a compensation for their time.

\subsection{Variables}
One independent variable was considered: \textbf{Score Attribution System} with two possible values $\{Self\_I, E\_Altr\}$.
Three dependent (within-subjects) variables were considered:
\begin{itemize}
\item
\textbf{Mean number of takes (per player)}, the mean number of letters a player acquired for himself/herself in each level. The value space is $[0,4]$, as each player could perform at most four actions (gives or takes) per level. This variable was measured by analyzing game logs;
\item
\textbf{Final game score (per player)}, the score obtained at the end of playing each version of the game. This variable was acknowledged to support the differences observed by the mean number of takes;
\item
\textbf{Motives orientation (per player)}, which measured the motive behind players' interactions, between self-oriented and others-oriented. This measure was obtained at the end of each played version, through to a question ``Who did I focus while playing this version?'', answered using a seven-point Likert scale ranging from ``Me'' to ``The other player'';
\end{itemize}


\section{Results}
\label{section:Results}

\subsection{Mean number of takes and Final score}
\begin{figure}
	\centering
	\includegraphics[width=0.8\linewidth]{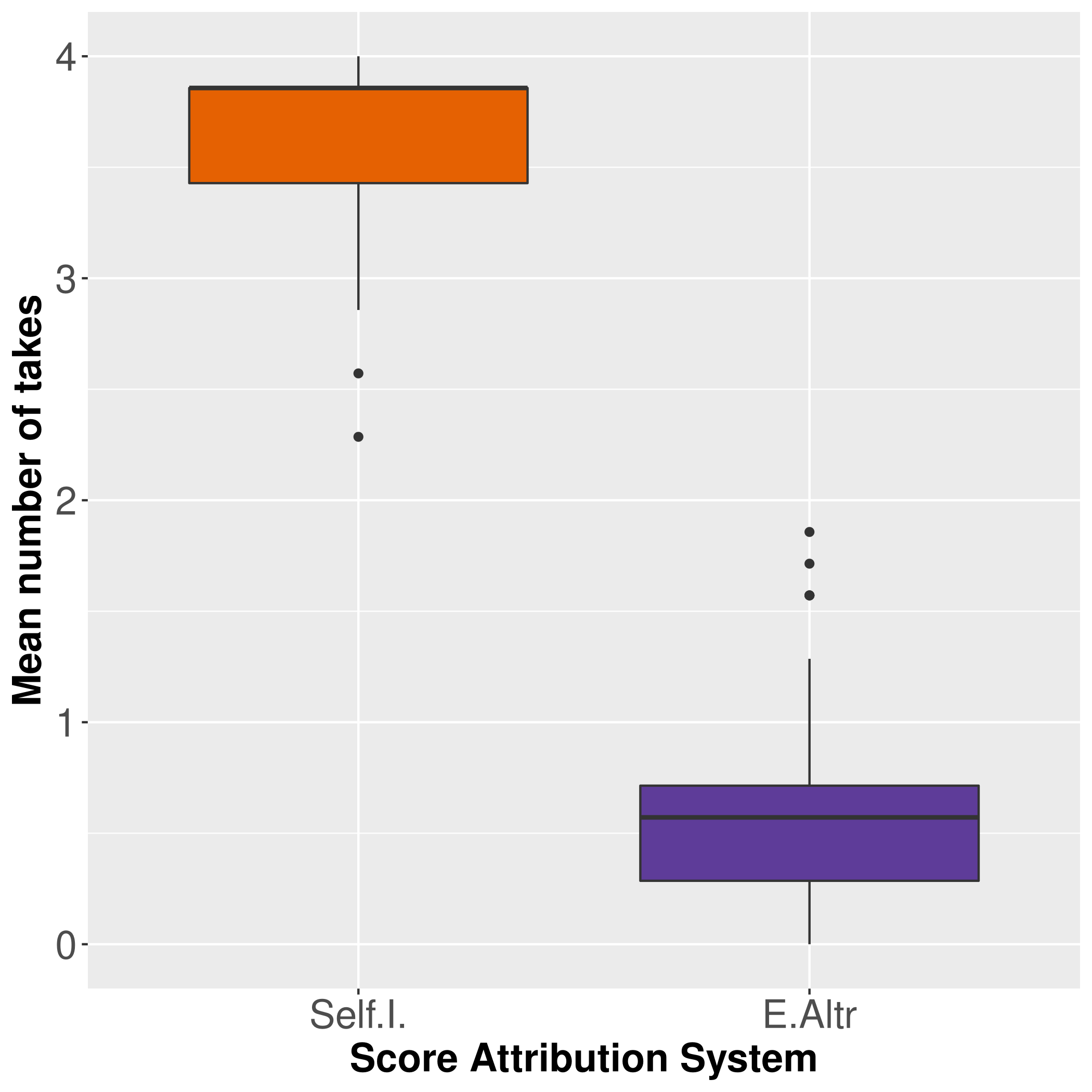}
	\caption{\label{fig:meanNumberOfTakes} Distribution of mean number of takes by score system.}
\end{figure}

\begin{figure}
	\centering
	\includegraphics[width=0.9\linewidth]{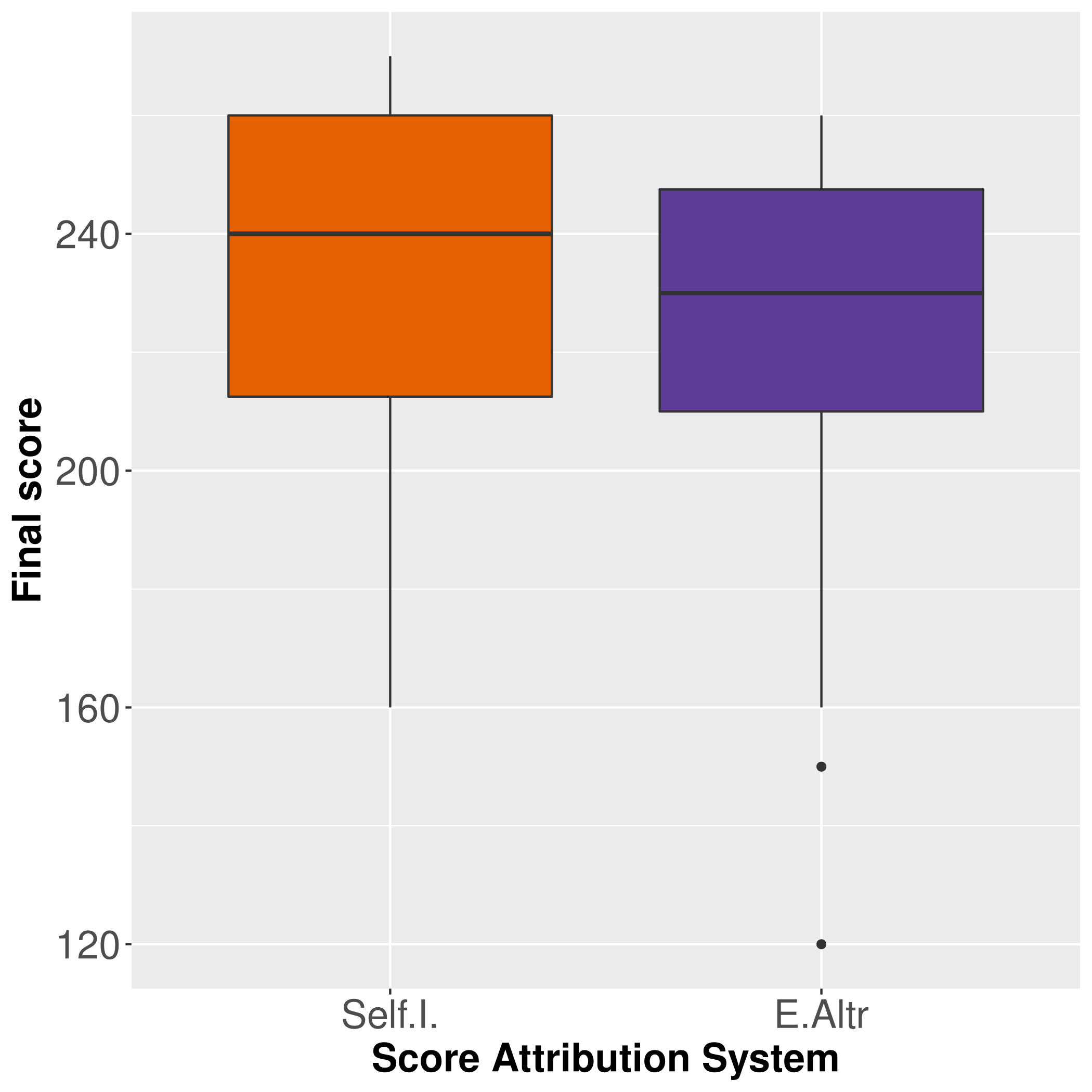}
	\caption{\label{fig:finalScore} Distribution of final scores by score system.}
\end{figure}

The distribution of mean number of takes is plotted in Fig. \ref{fig:meanNumberOfTakes}.
Shapiro-Wilk tests reported a non-normal distribution regarding the mean number of takes values.
Therefore, a Wilcoxon paired signed-rank test was performed to compare the two score systems, at the level of significance $p=0.05$. 
The number of performed take actions changed highly significantly between the two score strategies, with a large effect size ($Z = 7.06, p<<-.001, r = -.87$).
Furthermore, if we analyze the distribution of the data, we can observe that players of \textit{Self-Improvement} score systems performed, on average, a high (near maximum) number of takes $(M\approx3.64,  Mdn\approx3.86, SD\approx0.37)$, while oppositely, players of the \textit{Extremely Altruistic} score system performed a low number of takes $(M\approx-.58, Mdn\approx-.57, SD\approx0.42)$. 
Even though the differences are clearly noticeable, we can also observe that the Self-Improvement data is closer to the maximum number of takes, than the altruistic version is to the minimum number of takes. This effect was possibly caused by the fact that while searching for the most rewarding strategies, the players found easier or more natural to start by taking letters for themselves, which resulted in differences in the final scores (see Fig. \ref{fig:finalScore}).
\textit{In summary, these results support that players implicitly understood that the optimal strategy while playing the \textit{Self-Improvement} version was to take letters, and the optimal strategy while playing the \textit{Extremely Altruistic} version was to give letters, even though these strategies where unknown throughout the game execution.}


\subsection{Motives orientation}
\begin{figure}
	\centering
	\includegraphics[width=0.9\linewidth]{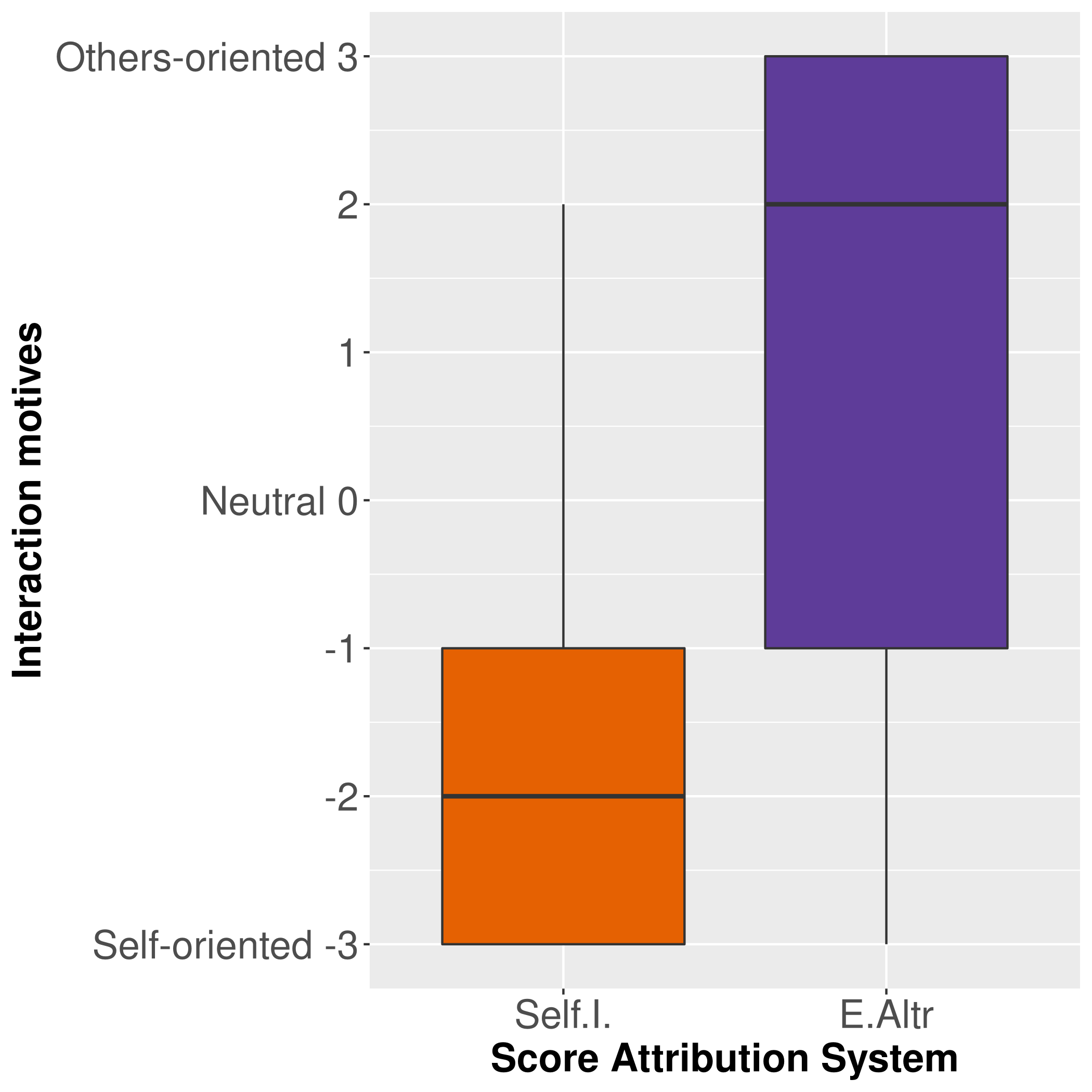}
	\caption{\label{fig:interactionMotives} Distribution of motive orientation by score system.}
\end{figure}

The distribution of the motive orientation values is plotted in Fig. \ref{fig:interactionMotives}.
Shapiro-Wilk tests reported a non-normal distribution regarding the motives orientation values.
Therefore, a Wilcoxon paired signed-rank test was conducted, at the level of significance $p=0.05$.  
The motive orientation values also varied highly significantly between the two score attribution strategies, with a large effect size ($Z = -5.73,  p<<-.001, r = -.71$).
By observing the distribution of the data, we can conclude that \textit{Self-Improvement} responses were driven towards ``Self-oriented'' $(M\approx-1.76, Mdn=-2, SD\approx1.43)$, opposite to \textit{Extreme Altruism} responses, which were approximated towards ``Others-oriented'' $(M\approx1.12, Mdn=2, SD\approx2.30)$.
\textit{These results indicate that, as expected, the \textit{Self-Improvement} version was perceived as allowing players to improve their gameplay, and the \textit{Extremely Altruistic} score system was perceived as a helping scenario implying that players' interactions were exclusively motivated by other players.}

\subsection{Discussion}
\label{section:Discussion}

In this study we developed and compared the effects of two reward parameterizations, representing two extreme behavior styles: \textit{Self-Improvement} and \textit{Extreme Altruism} in the players strategies and self-reported interaction motives. Notably, the players played through several game levels, without knowing how the game was scored.
Results indicated that the \textit{Self-Improvement} version led players to perform a high (near maximum) number of takes and report self-oriented interaction motives, while the \textit{Extreme Altruism} version led players to perform a low (near minimum) number of takes and report others-oriented interaction motives. Thus, our expectations were corroborated. Moreover, the tendencies revealed high effect sizes, which means strong differences were mediated by the two score functions.
It is important to note that others-oriented interactions were still achieved, even though a limit to the number of actions was imposed which could bias the strategies and interaction motives towards \textit{Self-Improvement}. It is also important to note that rich and dynamic interactions were promoted, even though there was the concern that players could possibly deviate their focus to the game tasks alone, without acknowledging their rewards. 
Furthermore, although players acquired high scores in both conditions, they still significantly acquired higher scores in Self-Improvement condition. This may indicate a natural tendency for players to start exploring the effects of their own actions, before exploring others-oriented actions. This corroborates the argument that interactions with individual motives might be more easily  discovered than others-motivated interactions \cite{correia2019exploring}. 
The aforementioned results reflect an important finding, which answers our hypotheses and research question: 
\textit{our score attribution strategies led all players to implicitly adopt significantly different in-game strategies (mean number of takes) and to report different interaction motive orientations for the two game versions. Thus, evidence was obtained for individual and altruistic reward mediated behavior}. 
In other words, without altering the game mechanics or awarding different game items, players still managed to learn meaningful strategies, which motivated them to interact in different ways, thus proving the effectiveness of our approach.

Besides exclusively examining behavior promotion, this study sheds light on the application of individual- or altruistic-oriented rewards to regulate and balance dilemmas in which the players' decisions might be influenced by individual differences and cultural backgrounds \cite{parks1994social, hagger2014evaluating}. 
Additionally, these techniques can also be used for behavior change. For instance, to promote altruistic attitudes through an environment conservation game, an altruistic inducing strategy can be applied in a public goods game which aims to alert people for environmental sustainability.

The further consideration of a similar strategy must, however, be treated with care, as in order to apply our strategy, our game presented tasks that could be completed by different means, but using the same player actions. This may not be the case of the tasks present in all games.
In more complex scenarios, instead of using a reward-based approach \textit{per se}, this technique can be used to complement other interaction mechanics in ways that might be further investigated in different serious games or in specific social dilemmas implementations.
\section{Conclusions}
\label{section:Conclusions}

In this work we approached the promotion of altruistic and individual behaviors exclusively through the use of rewards. 
Our premise was that an altruistic behavior, modeled by a full others-directed interaction motive orientation, could be implicitly incentivized by rewarding players who contributed to the completion of other players' tasks, while, oppositely, an individual behavior, modeled by a full self-directed interaction motive orientation, could be implicitly incentivized by rewarding players who contributed to the completion of their own tasks.
To test the validity of these assumptions, we deployed two different score attribution systems in a word-matching game named Message Across, and conducted several user tests, where participants did not know how the games were scored. The results indicated accentuated tendencies for the promotion of both altruistic and individual behavior, as highly significant differences, aligned to our expectations, were observed for the players' strategies and self-reported interaction motive orientations.
In particular, the self-oriented version drove players to perform a high number of takes and an altruism-oriented version drove players to perform a low number of takes. Besides, these diverging players' strategies allowed the self-oriented version to motivate players to focus on themselves, and the altruism-oriented version to motivate players to focus on other players.


Future research includes the investigation of whether individual differences such as personality have an effect on how people vary their playing strategies and interaction motives regarding individual and altruistic behavior.
We also believe that varying the size of words or using different numbers of shared letters is worthwhile to verify how the length and type of the task may have an impact on the players' focus and playing styles.
A cross cultural study would also be interesting, given the already presented tendency for culture to influence players' strategies and behavior evaluations \cite{parks1994social, hagger2014evaluating}.

Furthermore, we can extend the analysis of interaction motives and playing styles to interactions beyond the ones measured in the present study. In fact, we can analyze our interactions motives scale in higher granularity, by considering the less symmetric behaviors such as Mutual Help (also acknowledged by some of the presented related research of Section \ref{section:RelatedWorkI}) and Competition. 

Finally, our findings also contribute to the field of automatic education and training, due to the importance of behavior promotion for this research topic.
Models such as GIMME~\cite{gomes2019gimme}, that aim to optimize the collective ability of groups of people interacting with one another, may use scores to mediate students' interactions, thus empowering collective teaching in multiplayer settings.
Furthermore, promoting interaction styles using rewards allows researchers to take a more human approach to the integration of agents that simulate people in serious games, besides adding expressiveness to their simulation models.



\bibliographystyle{IEEEtran}
\bibliography{IEEEabrv,References}

\end{document}